\documentclass[a4paper]{jpconf}
\usepackage{graphicx}
\usepackage{amsmath}
\usepackage{cite}
\usepackage{color}

\begin{document}
\title{Microwave photon generation in a doubly tunable superconducting resonator}

\author{Ida-Maria Svensson$^1$ , Mathieu Pierre, Micha\"el Simoen, Waltraut~Wustmann, Philip Krantz, Andreas~Bengtsson, G\"oran~Johansson, Jonas Bylander, Vitaly Shumeiko and Per~Delsing$^2$}
\address{Microtechnology and Nanoscience, Chalmers University of Technology, SE-41296 G\"{o}teborg, Sweden}
\ead{$^1$ida-maria.svensson@chalmers.se}
\ead{$^2$per.delsing@chalmers.se}
%\address{Laboratoire National des Champs Magnétiques Intenses (LNCMI), Université de Toulouse INSA UPS, CNRS UPR 3228, EMFL, FR-31400 Toulouse, France}

\begin{abstract}
%Max 200 words.
We have developed and tested a doubly tunable resonator, with the intention to simulate fast motion of the resonator boundaries in real space. Our device is a superconducting coplanar-waveguide half-wavelength microwave resonator, with fundamental resonant frequency $\sim5\,$GHz. Both of its ends are terminated by dc-SQUIDs, which serve as magnetic-flux-controlled inductances. Applying a flux to either SQUID allows tuning of the resonant frequency by approximately $700\,$MHz. By using two separate on-chip magnetic-flux lines, we modulate the SQUIDs with two tones of equal frequency, close to twice that of the resonator's fundamental mode. We observe photon generation, at the fundamental frequency, above a certain pump amplitude threshold. By varying the relative phase of the two pumps we are able to control the photon generation threshold, in good agreement with a theoretical model for the modulation of the boundary conditions. At the same time, some of our observations deviate from the theoretical predictions, which we attribute to parasitic couplings, resulting in current driving of the SQUIDs.
\end{abstract}

%The new feature we have here is the two SQUIDs and double tunability.
%Left/Upper (SQUID \#1) and Right/lower (SQUID \#2)!
%Compare single SQUID $\lambda/2$, sample Nb38Schip57
%Breathing cavity sounds better than breathing resonator but doubly tunable resonator sounds much better than doubly tunable cavity. I have used the word resonator,this could be changed.

%\section{To check and do}
%\begin{itemize}
%	\item
%\end{itemize}

\section{Introduction}
Vacuum is commonly considered to be empty space. However, in quantum theory, vacuum is in fact not empty but contains vacuum fluctuations of the electromagnetic field. Due to these fluctuations, two perfectly conducting mirrors at rest, placed in close vicinity of each other, can exhibit radiation pressure forces, known as the Casimir effect \cite{Casimir1948}. Further, if the mirrors are moved with a speed close to the speed of light, real photons can be generated as excitations of the vacuum fluctuations, a phenomenon called the dynamical Casimir effect (DCE) \cite{Moore1970}. In fact, photon generation through the DCE does not require two mirrors: one fast moving mirror is enough to produce DCE photons \cite{DeWitt1975,Fulling1976}. 

Using superconducting circuits, the physical conditions equivalent to a mirror moving at $1/4$ of the speed of light can be created \cite{Sandberg2008}. This is done by placing a superconducting quantum interference device (SQUID) at the end of a transmission line. The SQUID acts as a tunable inductance, $L_J(\Phi_{ext},I_s)=\Phi_0/\left(2\pi |\cos(\Phi_{ext}\pi/\Phi_0)|\sqrt{I_c^2-I_s^2}\right)$, where $\Phi_0$ is the magnetic flux quantum, $\Phi_{ext}=\Phi_{dc}+\Phi_{ac}(t)$ is the applied external magnetic flux, $I_c$ is the SQUID's critical current, and $I_s$ the current through the SQUID. The SQUID inductance can be modulated either by flux pumping, through $\Phi_{ac}$, which is a direct modulation of the boundary condition of the resonator and the analogue of a moving mirror, or by ac driving of the SQUID current $I_s$. The generation of DCE photons, using a flux-pumped SQUID at the end of a transmission line was suggested in Ref. \cite{Johansson2010} and demonstrated in Ref. \cite{Wilson2011}.

If a SQUID is included in a resonator and flux-modulated around twice the resonant frequency, the system is the equivalent of a parametric oscillator \cite{Dykman1998,Wilson2010,Wustmann2013}, \textit{i.e.} a harmonic oscillator driven by the modulation of a system parameter, the resonant frequency. The parametric oscillator has a flux pump amplitude threshold, above which self-sustained oscillations are generated \cite{Wilson2010}, determined by the system damping. Below threshold, the system can be operated as a parametric amplifier in which small input signals near its resonant frequency are amplified \cite{Castellanos2007,Tholen2007,Yamamoto2008,Roch2012}.

In this paper we extend the concept of the tunable resonator to have one SQUID in each end, with the intention of forming two independently controllable boundary conditions. We use a $\lambda/2$ (half-wavelength) superconducting coplanar waveguide resonator, with each end terminated by a SQUID. If driven separately, both SQUIDs can generate photons individually through the DCE. When driven together at the same frequency, the resonator can be thought of as a vibrating resonator or a breathing resonator, depending on the phase difference between the two drive signals. When flux pumping both SQUIDs around twice the first resonator mode, theory predicts constructive interference for the breathing mode, leading to a low threshold for photon generation, and destructive interference for the vibrating mode, \textit{i.e.} no photon generation \cite{Lambrecht1996,Ji1998,Dodonov1998,Lambrecht1998,Dalvit1999}. In addition to investigations of the DCE this device opens up doors for future interesting experiments, for example, measurements of the twin paradox \cite{Lindkvist2014}, where a microwave signal could be sent on a ``space trip'' in a vibrating resonator, and generation of cluster states \cite{Bruschi2016}.

%We treat pumping around $2\omega_1$, which means that the effective strength of the pump is
%\begin{align}
%\epsilon_{eff}=\frac{\omega_1\cos^2(\omega_1d/2v)}{2M_1(\omega_1d/v)^2}\left[\frac{|\tan(\Phi_{dc,l}\pi/\Phi_0)|}{\gamma_l}\delta f_l+\frac{|\tan(\Phi_{dc,r}\pi/\Phi_0)|}{\gamma_r}\delta f_r\right],
%\end{align}
%where the pump signals $\delta f$ are complex numbers with an amplitude and a phase, $\delta f_{l/r}=(\Phi_{ac,l/r}\pi/\Phi_0)e^{i\varphi_{l/r}}$. For the first resonator mode we have the generalized mode mass $M_1=1+\sin(\omega_1d/v)/(\omega_1d/v)\approx 1.$
%Then the threshold for parametric oscillations with degenerate pumping $\omega_p\approx 2\omega_1$ is
%\begin{equation}
%\label{eq:thresh}
%\epsilon_{eff}=\epsilon_{thresh}=\sqrt{\Gamma^2+\delta^2},
%\end{equation}
%where $\Gamma$ is the photon loss rate of the resonator and $\delta=\omega_p/2-\omega_1$ is the detuning of the pump signal. For $\epsilon_{eff}>\epsilon_{thresh}$ and degenerate pumping, the number of produced photons in the resonator is $N_1=\left(-\delta+\sqrt{\epsilon_{eff}^2-\Gamma^2}\right)/\alpha_1$,
%where $\alpha_1$ is the Duffing parameter, which is set by the non-linearity in the SQUIDs as $\alpha_1=\alpha_{1,l}+\alpha_{1,r}=\hbar/(2E_{L,cav})\cos^4(\omega_1d/2v)/(M_1^2\omega_1^2(d/v)^4)\left(1/\gamma_l+1/\gamma_r\right).$ Here $E_{L,cav}=\left(\Phi_0/2\pi\right)^2 1/L_0d$ denotes the inductive energy of the resonator.

\section{Experimental setup}
%Present in principle sample Nb34Schip62 as a double SQUID resonator and NbXXSchipXX as a single SQUID resonator.
\begin{figure}
\begin{minipage}[c]{10.2cm}
\includegraphics{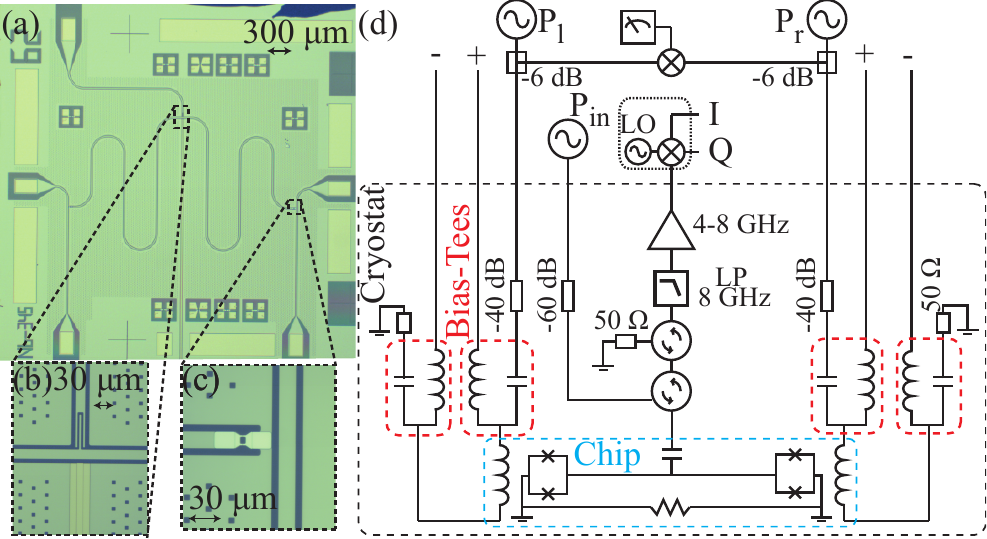}
\end{minipage}
\begin{minipage}[c]{5.8cm}
\caption{\label{fig:SampleAndSetup}{\small(a) Micrograph of the doubly tunable resonator chip. (b) The middle of the resonator with a gold-filled slot (below) and a coupling capacitor and probe (above). (c) End of the resonator with the SQUID and on-chip flux line. (d) Schematic of the measurement setup. It is a reflection setup with circulators to probe the resonant frequency and also to measure the output when the resonator is pumped through the on-chip flux lines.}}
\end{minipage}
\end{figure}
In Fig.~\ref{fig:SampleAndSetup} we present the sample layout and measurement setup. The (a), (b) and (c)-panels show micrographs of the sample. The SQUIDs are made of aluminium and deposited by two-angle evaporation, while the rest of the circuit is etched in niobium. Everything is placed on a sapphire substrate. The resonator is meandered and grounded in both ends. To avoid a parasitic superconducting loop through resonator and ground plane, we made a slot in the ground plane. To keep good electrical contact we bridged the slot with normal metal (gold), see Fig.~\ref{fig:SampleAndSetup}(b). 

The measurement setup is a reflection setup (Fig.~\ref{fig:SampleAndSetup}(d)) with circulators to allow for proper attenuation of the input signal and amplification of the output signal. The flux line setups enable both dc biasing and fast modulation. The dc and ac flux signals are combined in bias-Tees at the mixing chamber stage of the cryostat. The fast modulation signals are sent from two separate signal generators, $P_l$ and $P_r$, which are phase locked by a $10\,$MHz reference. To allow for measurement of the phase difference between the sources, the output signals are divided in power splitters and compared using a mixer. Provided that the two pump signals have the same frequency, the output of the mixer is a dc signal with varying amplitude, depending on the phase difference between the pump signals. The output signal from the resonator is down-converted and sampled in a digitizer, which records both the in-phase and out-of-phase quadrature.

\section{Measurement results - Resonator characterization}
\begin{figure}
\center
%\begin{minipage}[b]{13cm}
\includegraphics{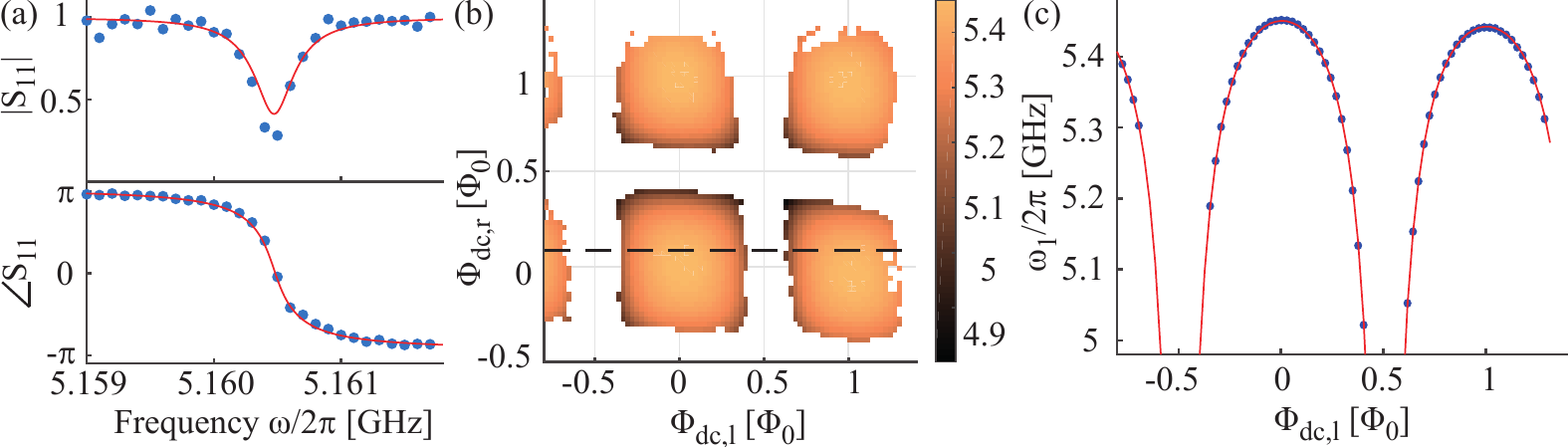}
%\end{minipage}
%\begin{minipage}[b]{3cm}
\caption{\label{fig:DCtuning}{\small(a) Reflection measurement (blue dots) at $\Phi_{dc}=(0.3,0.3)\,\Phi_0$ and a fit to the model $S_{11}=(1/Q_{ext}-1/Q_{int}-2i(\omega-\omega_1)/\omega_1)/(1/Q_{ext}+1/Q_{int}+2i(\omega-\omega_1)/\omega_1)$. For this bias point we can extract $\omega_1/2\pi=5.1605\,$GHz, $Q_{ext}=15.4\cdot 10^3$ and $Q_{int}=36.9\cdot 10^3$. (b) dc tuning of the resonant frequency by both magnetic flux biases. (c) Linecut from (b) indicated by the black dashed line. The blue dots are data and the red line is a fit to the model in Eq.~\eqref{eq:Spectrum}.}}
%\end{minipage}
\end{figure}

We can tune the resonant frequency by controlling the two dc-fluxes, $\Phi_{dc,l/r}$. The first resonator mode is probed by measuring the reflection coefficient of a microwave signal incident on the resonator (Fig.~\ref{fig:DCtuning}(a)); the extracted resonant frequencies are presented in Fig. \ref{fig:DCtuning}(b). The pattern is slightly tilted due to a small inductive crosstalk.

The second resonator mode is outside the frequency band of our amplifier and circulators, but we can determine the mode frequency using parametric up-conversion \cite{ZakkaBajjani2011,Wustmann2017}. This is a two-photon process where a weak probe signal is exciting the first mode, and at the same time one of the SQUIDs is flux-pumped at the difference frequency between the second and the first mode, $\omega_2-\omega_1$. When the pump hits the difference frequency, photons are converted between the modes, an avoided level crossing can be measured and the second mode frequency can be extracted. In Table \ref{tab:f2} we list three measured points. We can conclude that the anharmonicity is much larger than the linewidth of the resonator.

\begin{table}
	\begin{minipage}[c]{7.8cm}
		\caption{\label{tab:f2}{\small Two-tone spectroscopy measurements of the second resonator mode using parametric up-conversion. The first column indicates the flux bias point, $\omega_1$ and $\omega_2$ are the mode frequencies of the two lowest modes and the last column is the spectrum anharmonicity.}}
	\end{minipage}
	%\begin{center}
	\begin{minipage}[c]{8.2cm}
		\begin{tabular}{|cccc|}
			\hline
			$(\Phi_{dc,l},\Phi_{dc,r})$  & $\omega_1/2\pi$ & $\omega_2/2\pi$ & $(2\omega_1-\omega_2)/2\pi$ \\
			$[\Phi_0]$ & [GHz] & [GHz] & [MHz]\\
			\hline
			(0.01,0.01)& $5.459\,$ & $10.867\,$ & $47\,$  \\
			(0.21,-0.19)&  $5.360\,$ & $10.668\,$ & $52\,$  \\
			(0.31,-0.29)& $5.184\,$ & $10.323\,$ & $45\,$  \\ \hline
		\end{tabular}
	\end{minipage}
	%\end{center}
\end{table}

By straightforward extension of the results in Ref.~\cite{Wustmann2013}, we find that the spectrum of the doubly tunable resonator is described by the equation
\begin{align}
\frac{\omega_n}{v}d\tan\left( \frac{\omega_n}{v}d\right)\left[1- \left(\frac{v}{\omega_nd}\right)^2\left(\frac{1}{\gamma_l}-c\left(\frac{\omega_n}{v}d\right)^2\right)\left(\frac{1}{\gamma_r}-c\left(\frac{\omega_n}{v}d\right)^2\right)\right]=\frac{1}{\gamma_l}+\frac{1}{\gamma_r}-2c\left(\frac{\omega_n}{v}d\right)^2,
\label{eq:Spectrum}
\end{align}
where the subscripts $l/r$ correspond to the left and right SQUID, respectively, $\omega_n$ is the frequency of mode $n$, $d=10.133\,$mm is the resonator length and $v=1/\sqrt{C_0L_0}$ is the phase velocity. $\gamma_{l/r}=L_{J,l/r}/(L_0d)$ is the inductive participation ratio for each SQUID, where the SQUID inductance is $L_{J,l/r}=\Phi_0/(2\pi I_{c}|\cos(\Phi_{dc,l/r}\pi/\Phi_0)|)$ and $\Phi_{dc,l/r}$ is the static magnetic flux bias of the left and right SQUID respectively, assuming low signal levels, $I_s\ll I_c$. The capacitive participation ratio, $c=C_{J}/C_0d$, where $C_{J}$ is the SQUID capacitance, is much smaller than the inductive contribution, grows with the mode number $n$. Here we have assumed that the two SQUIDs are nominally identical, $\gamma_0=\gamma_{0,l}=\gamma_{0,r}$ and $C_J=C_{J,l}=C_{J,r}$.

\begin{table}
\caption{\label{tab:DCfit}{\small Extracted parameters for the resonator. The inductive participation ratio is $\gamma_0=L_{J,0}/L_0d$, $I_c$ the SQUID critical current, $C_J$ the SQUID capacitance and $\xi_{l/r}$ are the dc-crosstalk parameters. $C_0$ and $L_0$ are the capacitance and inductance per unit length of the coplanar waveguide, respectively. $\omega_1$ is the resonant frequency of the lowest mode, $\Gamma$ is the photon loss rate and $Q_{\rm{int}}$ and $Q_{\rm{ext}}$ are the quality factors of the resonator at $\Phi_{dc}=(0,0)\,\Phi_0$. The translation between the loss rate $\Gamma$ and the Q-values is $2\Gamma=\omega_1/Q_{\rm{int}}+\omega_1/Q_{\rm{ext}}$.}}
\begin{center}
\begin{tabular}{|ccccccccccc|}
\hline
$\gamma_0$  & $I_c$ & $C_J$ & $\xi_l$  & $\xi_r $  & $C_0$  & $L_0$  & $\omega_1$  & $2\Gamma$  & $Q_{\rm{int}}$ & $Q_{\rm{ext}}$\\
& & & & & & & $(\Phi_{dc}=0)$ & $(\Phi_{dc}=0)$ & $(\Phi_{dc}=0)$ & $(\Phi_{dc}=0)$ \\
$[\%]$  & [$\mu$A] & [fF] & $[\%]$  & $[\%]$  & $[\frac{\rm n \rm F}{\rm m}]$  & $[\frac{\rm \mu \rm H}{\rm m}]$  & [GHz]  & [MHz]  & [$10^3$] & [$10^3$]\\ \hline
$4.64$  & 1.64 & 89 & $3.64$  & $4.19$  & $0.159$  & $0.427$  & 5.459  & $0.56$  & 400 & 9.6\\ \hline
\end{tabular}
\end{center}
\end{table}

The two-dimensional dc tuning, Fig.~\ref{fig:DCtuning}(b), together with the measurements of the second mode shown in Table \ref{tab:f2}, can be fitted using Eq.~\eqref{eq:Spectrum}. A linecut of Fig.~\ref{fig:DCtuning}(b) with a fit is displayed in Fig.~\ref{fig:DCtuning}(c). Extracted resonator and SQUID parameters are presented in Table~\ref{tab:DCfit}. The $\xi_{l/r}$ parameters are the dc crosstalks, \textit{i.e.} how much each SQUID is affected by the opposite flux line (only a few percent of the coupling from the closest flux line). The resonant frequency can be tuned over a wide frequency range, where the limiting factor is the photon loss rate. The photon loss rate, which increases as $\Phi_{dc,l/r}$ approaches $\Phi_0/2$. However, as seen in Fig. \ref{fig:DCtuning}(b), resonant frequencies below $4.9\,$GHz are measurable. 

\section{Measurement results - Pumping}
\begin{figure}
	\begin{center}
		\includegraphics{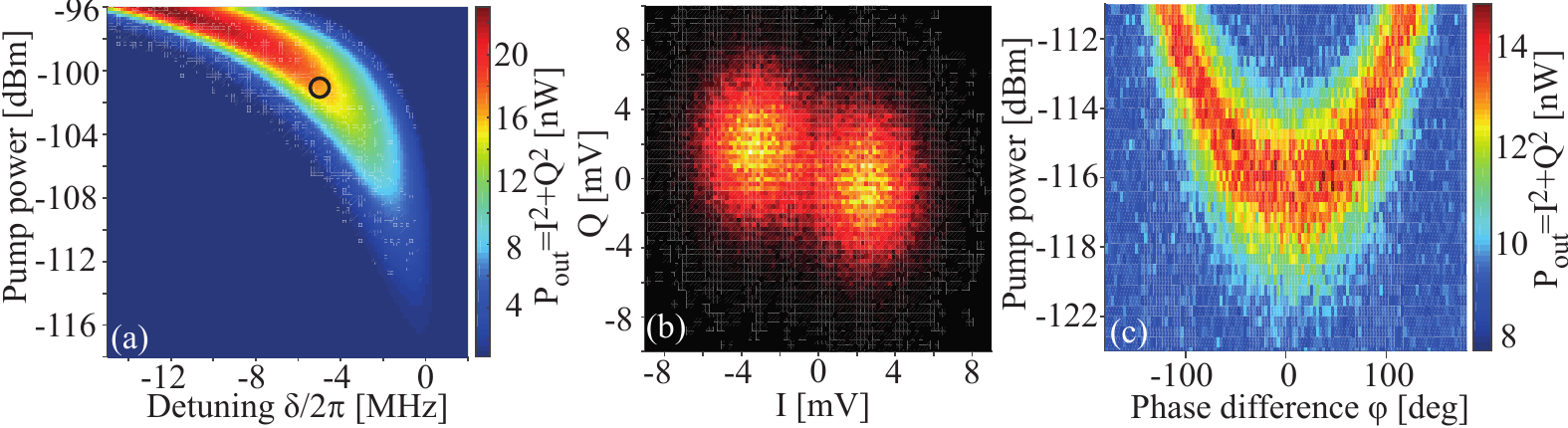}
		\caption{\label{fig:Pumping}{\small (a) Photon down-conversion. Measured with a single pump applied to the left flux line at the bias point $(0.3,0.3)\,\Phi_0$. (b) Histogram taken at the point marked with a black circle in (a). We measure two $\pi$-shifted states. (c) Double-pump measurement, where the phase difference, $\varphi$ between the pump signals is varied. Here the SQUID bias is $(0.2,0.2)\,\Phi_0$ and the generated radiation for $\delta=-1\,$MHz is displayed.}}
	\end{center}
\end{figure}

%K2-upper SQUID and K3-lower SQUID
By applying a pump tone to one of the ac flux lines at a frequency close to $2\omega_1$, we expect to observe parametric oscillations. We measure the quadrature components of the output signal, and calculate the total output power, $P_{out}=\langle I^2\rangle+\langle Q^2\rangle$. Fig.~\ref{fig:Pumping}(a) shows photon down-conversion in a range of detuning and pump power. The detuning is denoted $\delta=\omega_p/2-\omega_1$, where $\omega_p$ is the pump frequency. Furthermore, we sample the individual quadratures, $\langle I(t)\rangle$ and $\langle Q(t)\rangle$, and histogram $1\cdot 10^5$ samples, see Fig. \ref{fig:Pumping}(b). The histogram shows two stable $\pi$-shifted states with the same amplitude, characteristic for parametric oscillations \cite{Dykman1998,Wilson2010,Krantz2016}.

We can also apply pump signals to both flux lines simultaneously. The amplitudes are adjusted such that the effective pump strengths of the two individual SQUIDs are equal. This was done by measuring single-pump thresholds, which for the bias point $(0.2,0.2)\,\Phi_0$ should be equal. We find that, depending on the phase difference $\varphi=\varphi_r-\varphi_l$, the threshold for photon generation changes, see Fig.~\ref{fig:Pumping}(c). 

The theoretical prediction of the parametric oscillation threshold is $\epsilon_{th} =\sqrt{\Gamma^2 + \delta^2}$, which is the same for both the single and double pump case. This formula is symmetric in $\delta$, which our measured oscillation regions are not, due to a pump-induced frequency shift, because of the resonator nonlinearity, that shifts the resonant frequency towards red detuning. The threshold is reached when the effective pump strength $\epsilon_{eff}=\epsilon_{th}$. We follow the formalism \cite{Wustmann2013} and extend the results for the single pump to the double pump case. The effective pump strength is then a superposition of complex amplitudes of flux modulation in the left and right SQUIDs, $\Phi_{ac,l/r} = | \Phi_{ac,l/r}| e^{i\varphi_{l/r}}$, so that 
$\epsilon_{eff} = A(\omega_1)(k_l\Phi_{ac,l} + k_r\Phi_{ac,r}$). The coefficients of this superposition are, $k_{l/r}=|\tan(\Phi_{dc,l/r}\pi/\Phi_0)|/\gamma_{l/r}$. This gives an expected minimum threshold and therefore maximum photon generation in the breathing mode, $\varphi=0^\circ$, but cancellation and consequently no photons in the vibrating mode, $\varphi=\pm 180^\circ$. In agreement with measurement data.

\section{Discussion}
\begin{figure}
	\begin{center}
		\includegraphics{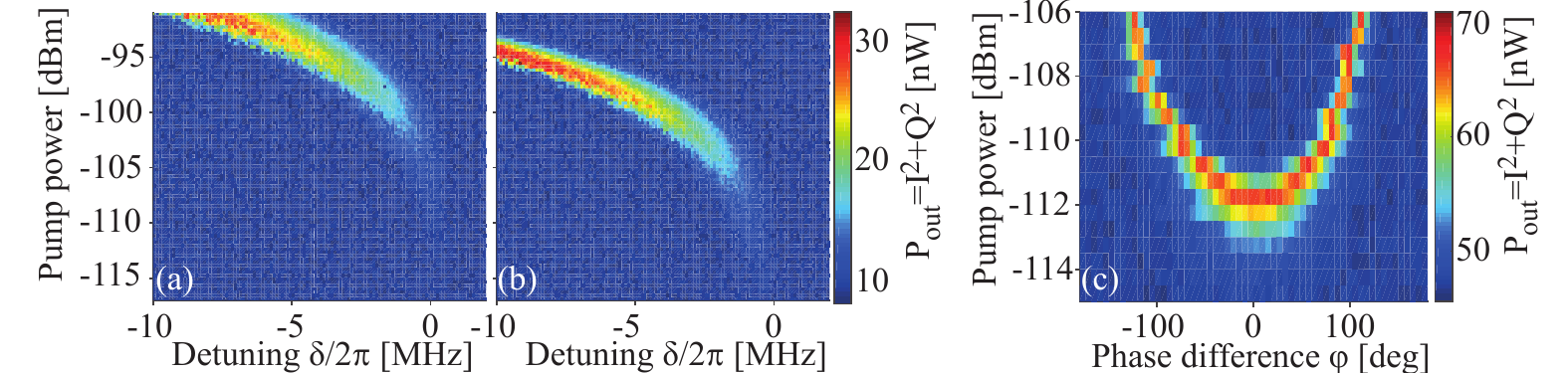}
		\caption{\label{fig:Contra}{\small Measurement results at SQUID bias $(0,0.2)\,\Phi_0$, in both cases using a single pump, coupled closest to the left (a), and the right (b) flux line respectively. (c) Double-pump measurement of a $\lambda/2$-resonator with only one SQUID, biased at $0.18\,\Phi_0$. Here $\delta=-6\,$MHz.}}
	\end{center}
\end{figure}

The results of Fig.~\ref{fig:Pumping} seem to agree qualitatively with the theory for a doubly flux-pumped resonator. However, we also observe some interesting deviations. In Fig.~\ref{fig:Contra}(a) and (b), we present regions of photon down-conversion, at the bias point $(0,0.2)\,\Phi_0$. The pump tone is applied to the left flux line in (a) and to the right in (b). Since for (a) the pumping is around $0\,\Phi_0$ and in (b) around $0.2\,\Phi_0$, we would expect differing results in the two graphs. However, the shapes of the oscillation regions in the two graphs are rather similar, although the thresholds differ by around $5\,$dB. It is surprising that we observe parametric oscillations at zero flux bias, since this contradicts theoretical predictions \cite{Wustmann2013}. We attribute this effect to a possible strong inductive ac crosstalk or a parasitic coupling. Even though we characterized the crosstalk at dc and found it negligible, it could be large at microwave frequencies, due to differences in signal distribution on the chip for dc and microwave signals. The observed $5\,$dB difference would correspond to $56\,\%$ ac crosstalk. Differences in setup attenuation cannot explain this large number.

A parasitic coupling from the flux pump to the SQUID current could occur, due to the presence of the low impedance loop through the resonator center conductor and the ground plane. This loop is $\sim 4000$ times larger than the SQUID loop, which corresponds to a significantly larger inductance. A coupling to this loop could cause circulating currents, and thereby direct driving of the SQUID current. A possible solution, making the loop less parasitic, would be to increase its impedance by changing the geometry of the gold-bridge and slot.

Another issue is the threshold pump strength. In experiments with a $\lambda/4$-resonator with identical SQUID flux-line design and similar resonant frequency, the single pump threshold is at least $20\,$dB higher than what is measured here. The difference in length of a $\lambda/2$ and $\lambda/4$-resonator could account for at the most a few dB of difference. Therefore the differing thresholds have to be explained, either by differing pumping mechanisms or significantly differing coupling between the flux line and SQUID. However, the latter can be ruled out since the coupling is designed to be identical.  

To find an explanation to the mentioned discrepancies, we performed a control experiment to probe the ac crosstalk. A similar resonator was fabricated with only one SQUID, \textit{i.e.}, the other end shorted to ground. Both resonator ends were equipped with on-chip flux lines, to allow for double-pump experiments. Surprisingly, we observe the same qualitative behaviour, independently of whether the resonator has two (Fig.~\ref{fig:Pumping}(c)) or one (Fig.~\ref{fig:Contra}(c)) SQUID. There are some differences in output power and oscillation region widths, but this is because the measurements were performed with different samples, in different setups, and at different bias points and detunings. This observation suggests an additional mechanism of down-conversion, possibly related to the microwave field filling the cavity and producing a current-pumping effect \cite{Yurke1988} as utilized in many parametric amplifiers \cite{Siddiqi2005,Castellanos2007}. The difference between flux and current pumping has been discussed in Ref. \cite{Krantz2016}. The phase dependence of the threshold in Fig.~\ref{fig:Contra}(c), could, for instance, be explained by direct interference by the two pump signals.

%There are some possible improvements that could cut the possibility of direct driving of the SQUID current. If one would use a stepped-impedance resonator, the mode frequencies could be engineered to be better separated from the pump frequencies. 

\section{Conclusion}
Using a $\lambda/2$ resonator with two magnetic-flux-tunable boundary conditions, we demonstrated photon generation by degenerate downconversion of a pump tone. When pumping with two signals at the same frequency, we observed a pump-phase dependence of the instability threshold for photon generation. This is in agreement with a theoretical model for modulation of the boundary conditions.
We also observed non-ideal results attributable to ac crosstalk and parasitic couplings resulting in current driving of the SQUIDs. 

\section*{Acknowledgements}
The authors acknowledge financial support from the European Research Council, the European project PROMISCE, Swedish Research Council and the Wallenberg Foundation. J.B. acknowledges partial support by the EU under REA grant agreement no. CIG-618353.

\section*{References}
\bibliography{refs}{}
\bibliographystyle{iopart-num}

\end{document}